\documentclass[conference]{IEEEtran}
\IEEEoverridecommandlockouts
\usepackage{cite}
\usepackage{amsmath,amssymb,amsfonts}
\usepackage{algorithmic}
\usepackage{hyperref}
\usepackage{graphicx}
\usepackage{textcomp}
\usepackage{subfigure}
\usepackage[T1]{fontenc}
\usepackage{mathrsfs}
\usepackage{multirow}
\usepackage{mathtools}
\usepackage{caption}

\usepackage[table, x11names]{xcolor}
\def\BibTeX{{\rm B\kern-.05em{\sc i\kern-.025em b}\kern-.08em
    T\kern-.1667em\lower.7ex\hbox{E}\kern-.125emX}}
    
\def\eg{\textit{e.g}. } 
\def\ie{\textit{i.e}. }

 \def\etal{\textit{et al}. }

\begin{document}

\title{Multi-Biometric Fuzzy Vault based on\\ Face and Fingerprints}

\author{\IEEEauthorblockN{C. Rathgeb$^*$, B. Tams$^*$, J. Merkle$^*$, V. Nesterowicz$^*$, U. Korte$^\dagger$, M. Neu$^\dagger$}
\IEEEauthorblockA{$^*$secunet Security Networks AG, Essen, Germany\\
$^\dagger$Federal Office for Information Security, Bonn, Germany   \\
\texttt{\{christian.rathgeb,benjamin.tams\}@secunet.com}}
}

\maketitle

\begin{abstract}
The fuzzy vault scheme has been established as cryptographic primitive suitable for privacy-preserving biometric authentication. To improve accuracy and privacy protection, biometric information of multiple characteristics can be fused at feature level prior to locking it in a fuzzy vault. We construct a multi-biometric fuzzy vault based on face and multiple fingerprints. On a multi-biometric database constructed from the FRGCv2 face  and the MCYT-100 fingerprint databases, a perfect recognition accuracy is achieved at a false accept security above 30 bits. Further, we provide a formalisation of feature-level fusion in multi-biometric fuzzy vaults, on the basis of which relevant security issues are elaborated. Said security issues, for which we define  countermeasures, are commonly ignored and may impair the overall system's security.
\end{abstract}

\section{Introduction}

\emph{Biometric template protection} \cite{Rathgeb11e,BNandakumar15a} refers to techniques that are capable of protecting biometric reference data (templates). These schemes obtain protected biometric templates, \ie pseudonymous identifiers (and auxiliary data), from unprotected biometric data. Biometric comparisons are then performed via protected biometric templates while unprotected biometric data are discarded. A general framework for biometric template protection methods is defined in ISO/IEC 24745 \cite{ISO11-TemplateProtection}. Said standard also stipulates the following properties of biometric template protection:

\begin{itemize}
\item \textit{Irreversibility}: the infeasibility of reconstructing the original biometric data given a protected template and its corresponding auxiliary data.  To achieve this goal a irreversible function is applied to extract the protected template. With this property fulfilled, the privacy protection is enhanced, and additionally the security of the system is increased against reconstruction attacks.
\item \textit{Renewability}: the possibility of revoking old protected templates and creating new ones from the same biometric instance and/or sample, \eg face image. With this property fulfilled, it is possible to revoke and re-issue the templates in case the database is compromised, thereby preventing misuse.
\item \textit{Unlinkability}: the infeasibility of determining if two or more protected templates were derived from the same biometric instance, \eg face. By fulfilling this property, cross-matching across different databases is prevented. This property can be achieved by incorporating a random secret. 
\item \textit{Performance preservation}: the requirement of the biometric performance not degrading the biometric authentication performance beyond the application-accepted tolerance.
\end{itemize}

\emph{Biometric cryptosystems} represent one category of biometric template protection methods. The majority of biometric cryptosystems binds a key to a biometric feature vector resulting in a protected template. Biometric comparison is then performed indirectly by verifying the correctness of a retrieved key \cite{BUludag04a}. Prominent examples of biometric cryptosystems are the fuzzy commitment \cite{BJuels99a} and fuzzy vault scheme \cite{BJuels02a}. The latter scheme was introduced by Juels and Sudan \cite{BJuels02a,bib:JuelsSudan2006} and enables protection and error-tolerant verification with unordered feature sets. It was firstly suggested for the protection of fingerprint minutiae sets in \cite{bib:ClancyKiyavashLin2003}, followed by a series of implementations of fingerprint fuzzy vaults \cite{bib:NandakumarJainPankanti2007,bib:Nagar2010}. So far, the fuzzy vault scheme has been successfully applied to different biometric characteristics, \eg iris \cite{BLee07b} and face \cite{RATHGEB2022102539}. 

It is well-known that a single biometric characteristic, \eg a single fingerprint or face, contains an insufficient amount of effective entropy to resist attacks exploiting the feature distribution, specifically false accept attacks \cite{Adler09,Lim16,Gong17}. Therefore, several researchers have proposed multi-biometric template protection systems \cite{Rathgeb12x}, \eg in \cite{Nagar12a}, which utilise biometric templates obtained from multiple characteristics or multiple instances of a single characteristic. In this context, multi-biometric fuzzy vault schemes have also been proposed, \eg in \cite{Nandakumar08,bib:Tams2015}.  In contrast to unprotected biometric systems, where fusion may also take place at score or decision level, the fusion in multi-biometric fuzzy vaults (and template protection systems in general) should be performed at the feature level to achieve high security levels \cite{Merkle12a}; otherwise, the protected templates of the individual characteristics can be attacked separately resulting in essentially the same security as of a fuzzy vault based on a single characteristic.

\setlength{\tabcolsep}{2.5pt}
\begin{table*}[!t]
\begin{center}
\caption{Most relevant approaches for multi-biometric fuzzy vault schemes based on a single and different biometric characteristics.}\label{tab:related}\vspace{-0.2cm}
\resizebox{\textwidth}{!}{
\renewcommand*{\arraystretch}{1.2}
\begin{tabular}{|c|c|c|c|c|c|c|}
\hline
\rowcolor{gray!15}\textbf{Approach} & \textbf{Year} & \textbf{Biometric Characteristics} & \textbf{Database(s)} & \textbf{Subjects} & \textbf{Performance Rates} & \textbf{Security Rates} \\\hline
\begin{tabular}{@{}c@{}}Nandakumar and\\ Jain \cite{Nandakumar08} \end{tabular}&  \begin{tabular}{@{}c@{}} 2008 \end{tabular} & \begin{tabular}{@{}c@{}}Fingerprint and iris\end{tabular} & \begin{tabular}{@{}c@{}} MSU DBI fingerprint,\\ CASIAv1 iris   \end{tabular} & 108 & GMR=98.2\% &  \begin{tabular}{@{}c@{}}49 bits\\ min-entropy \end{tabular}  \\\hline
\begin{tabular}{@{}c@{}}Meenakshi and \\Padmavathi  \cite{MEENAKSHI2010195} \end{tabular}&  \begin{tabular}{@{}c@{}} 2010 \end{tabular} & \begin{tabular}{@{}c@{}}Fingerprint, iris and retina\end{tabular} & \begin{tabular}{@{}c@{}} FVC02 DB2 fingerprint, \\CUHK iris, DRIVE retina  \end{tabular} & n.a. & n.a. &  \begin{tabular}{@{}c@{}}67 bits\\ min-entropy \end{tabular}  \\\hline
\begin{tabular}{@{}c@{}}Nagar \etal \cite{Nagar12a} \end{tabular}&  \begin{tabular}{@{}c@{}} 2012 \end{tabular} & \begin{tabular}{@{}c@{}}Face, fingerprint and iris\end{tabular} & \begin{tabular}{@{}c@{}}  FVC02 DB2 fingerprint,\\ CASIAv1 iris, XM2VTS face \end{tabular} & 100 & GMR=99\% &  \begin{tabular}{@{}c@{}}53 bits FAS \end{tabular}  \\\hline
\begin{tabular}{@{}c@{}}Tams \cite{Tams16a} \end{tabular}&  \begin{tabular}{@{}c@{}} 2016 \end{tabular} & \begin{tabular}{@{}c@{}}Four fingerprints\end{tabular} & \begin{tabular}{@{}c@{}}  MCYT-100 \end{tabular} & 100 & GMR=95\% &  \begin{tabular}{@{}c@{}}60 bits FAS \end{tabular}  \\\hline
\begin{tabular}{@{}c@{}}Rathgeb \etal \cite{Rathgeb-ImprovedMultiBiometrics-EURASIP-2016} \end{tabular}&  \begin{tabular}{@{}c@{}} 2016 \end{tabular} & \begin{tabular}{@{}c@{}}Two irises\end{tabular} & \begin{tabular}{@{}c@{}}  IITDv1  \end{tabular} & 224  & GMR=95.5\% &  \begin{tabular}{@{}c@{}}42 bits FAS \end{tabular}  \\\hline
\end{tabular}
}
\end{center}
\hspace{0.15cm} GMR: Genuine Match Rate, FAS: False-Accept Security
\end{table*}

Works on multi-biometric fuzzy vaults commonly report an increase in biometric performance. However, they usually ignore potential security issues resulting from the fusion process. In particular, the nature of biometric templates extracted from different biometric characteristics may vary. More precisely, large differences in size and error distributions of the biometric templates can impair the security of a fuzzy vault scheme. In this work, we present a multi-biometric fuzzy vault based on the face and four fingerprints, which achieves auspicious biometric performance. To do so, a formalisation of feature-level fusion in multi-biometric fuzzy vault schemes is provided along with a detailed discussion of security aspects and theoretical definitions of countermeasures.  Eventually, we elaborate on the observed trade-off between biometric performance and security.

This work is organised as follows: related works and the fuzzy vault scheme are discussed in section~\ref{sec:related}  and \ref{sec:fvs}, respectively. Multi-biometric fuzzy vaults are formally detailed in section~\ref{sec:considerations}. Subsequently, a multi-biometric fuzzy vault based on face and fingerprints is presented in section~\ref{sec:casestudy}. A discussion is given in section~\ref{sec:discussion} and conclusions are summarised in section~\ref{sec:conclusion}.

\section{Related Works}\label{sec:related}
Table~\ref{tab:related} list the most relevant  scientific works on multi-biometric fuzzy vaults. In 2008, the first multi-biometric fuzzy vault scheme based on a feature fusion of fingerprint and iris was proposed by Nandakumar and Jain \cite{Nandakumar08}. In their scheme, the iris features are represented as fixed-length binary vector which cannot be protected directly in the fuzzy vault. To overcome this, the authors propose to transform the binary iris features based on a secret, \ie a transformation key. The transformation basically consists of Exclusive-OR (XOR) operations. The values of said transformation key are then locked in the fingerprint-based fuzzy vault while the transformed iris template is stored as additional auxiliary data. At the time of authentication, the transformation values are partially reconstructed by XORing a given iris vector with the auxiliary data. The reconstructed transformation values are subsequently used as unlocking set of the fuzzy vault.

Meenakshi and Padmavathi \cite{MEENAKSHI2010195} investigated the security of a fuzzy vault scheme based on features obtained from fingerprint, iris and retina. Features obtained from these characteristics are fused at feature level and are optionally hardened with a password. The password can be employed to transform the features prior to locking them in the vault. Unfortunately, the authors did not perform any biometric performance analysis. For fingerprints, minutiae points are extracted. On iris and retina images, basic image processing operators are applied to detect edges from which keypoints are extracted that are subsequently lock in the fuzzy vault.

Nagar \etal \cite{Nagar12a} propose a general framework for multi-biometric template protection based on feature level fusion. Within their framework, they apply different types of feature transformations to transform biometric feature vectors such that these can be locked in a fuzzy commitment or fuzzy vault scheme. With respect to the fuzzy vault scheme, Nagar \etal  fuse transformed features obtained from a single fingerprint, the face, and a single iris. Moreover, the authors highlight the issue that an attacker may succeed at unlocking a multi-biometric fuzzy vault by launching a false accept attack on a subset of used biometric characteristics. As countermeasure, they introduce a two-stage system in which fuzzy vaults based on each single employed biometric characteristic are constructed using unique keys. Subsequently, it is suggested to transform the original feature vectors by applying an irreversible function. Based on the transformed features, a multi-biometric fuzzy vault is constructed. The chosen key is used to further encrypt the fuzzy vaults based on single biometric characteristics using a conventional symmetric cipher. This means, at authentication, the multi-biometric fuzzy vault needs to be unlocked first. Afterwards, the recovered key will be used to decrypt the fuzzy vaults based on single characteristics. Finally, all fuzzy vaults based on a single characteristic need to be unlocked to recover all keys. However, the mentioned irreversible function is not specified and it is doubtable whether such a function exists. Note that such a function would require to guarantee irreversibility and at the same time should preserve the distances of features in the transformed domain.

The previously described works applied the fuzzy vault scheme to a fusion of features obtained from different biometric characteristics. In contrast, some researchers have also explored the use of multiple instances of a single biometric characteristic, \eg multiple fingerprints in \cite{Tams16a} or two irises in \cite{Rathgeb-ImprovedMultiBiometrics-EURASIP-2016}. These schemes have the advantage that the features obtained from different instances are processed in the same way and that a fusion at feature level can be performed through trivial concatenation.

Annapurani \etal \cite{Annapurani14} propose a multi-biometric fuzzy vault scheme based on fingerprint and ear using feature level fusion. They extract various types of features from both biometric characteristics. However, the authors perform their performance evaluations on a very small database consisting of 30 subjects. Moreover, security analyses are omitted.

Sujitha and Chitra \cite{Sujitha19} presented a multi-biometric fuzzy vault based on fingerprint and palmprint. In their system, feature vectors extracted from fingerprints are on average almost twice as long as those extracted from palmprints. The authors report a high security level while the obtained FMR indicates a significantly lower effective security. Unfortunately, the authors omit information about the used database(s). A similar approach based on fingerprint and palmprint was proposed by Brindha and Natarajan \cite{Brindha12}. However, the reported FMRs ($>$10\%) result in a very low security level.

A multi-biometric fuzzy vault based on fingerprint and finger vein was proposed by You and Wang \cite{You19}. Experiments were conducted on an in-house database and the number of subjects and used biometric samples was omitted. Further, the obtained security was not clearly reported in terms of bits.

In summary, previous works have shown that biometric performance can be significantly improved in a multi-biometric fuzzy vault works (compared to fuzzy vaults based on a single biometric instance). The majority of these works have suggested a fusion (concatenation) at feature level. However, they mostly ignore important security issues in multi-biometric fuzzy vaults resulting from the differences in the sizes and amount of noise of the fused feature sets that will be described in detail in the subsequent section. Finally, it should be noted that in the vast majority of published works, the security of multi-biometric fuzzy vaults is overestimated due to too optimistic assumptions.

\section{The Fuzzy Vault Scheme}\label{sec:fvs}

As mentioned before, the original fuzzy vault scheme was introduced in \cite{BJuels02a,bib:JuelsSudan2006}. It represents a cryptographic primitive for error-tolerant protection of unordered feature sets. Due to this ability, it has frequently been applied to secure fuzzy biometric features, in particular sets of minutae points. However, Security analyses have found that the original fuzzy vault scheme is vulnerable to linkage attacks \cite{bib:ScheirerBoult2007,bib:KholmatovYanikoglu2008}. In response to this issue, Dodis \etal~\cite{bib:DodisEtAl2008} presented an improved version of the fuzzy vault scheme that can effectively prevent from the aforementioned linkage attack \cite{bib:TamsMihailescuMunk2015}.

In the fuzzy vault scheme, an integer set $\mathbf{A}$ is interpreted as elements of a finite field, $\mathbf{P}\subset\mathbf{F}$. A secret polynomial $\kappa \in \mathbf{F}[X]$  of degree smaller than $k$ is generated (uniformly at random) and bound to the feature set $\mathbf{P}$ by computing the polynomial $V(X)=\kappa(X)+\prod_{a\in\mathbf{P}}  (X-a)$ as an instance of the improved fuzzy vault scheme. As final step of the enrolment, a cryptographic hash $H(\kappa)$ is stored along with $V(X)$ such that the pair $(V(X), H(\kappa))$ is considered as the vault record.

At verification, $\mathbf{U}\subset\mathbf{F}$ is obtained from an integer set $\mathbf{B}$. By evaluating the polynomial $V(X)$ on its elements, a set of pairs $\{(b, V(b)) | b \in  \mathbf{U}\}$ is obtained. The set $\mathbf{U}$ contains $\omega = |\mathbf{P}  \cap \mathbf{U}|$ genuine pairs lying on the function curve of the secret polynomial $\kappa(X)$. If the number $\omega$ of genuine points is at least $k$, it is possible to reconstruct the polynomial $\kappa$ from $\mathbf{U}$. The correctness of $\kappa$ can be verified by using the hash value $H(\kappa)$.

The polynomial reconstruction can be performed based on different algorithms. A Reed-Solomon decoder \cite{bib:Gao2002} has been suggested in the original fuzzy vault scheme \cite{BJuels02a,bib:JuelsSudan2006}. In \cite{bib:NandakumarJainPankanti2007}, the repeated use of a Lagrange-based decoder has been suggested. Moreover, a Guruswami-Sudan-based decoder \cite{bib:GuruswamiSudan1998} has been employed in various fuzzy vault schemes, \eg in \cite{Tams16a,RATHGEB2022102539}.

Finally, it is important to note that typical representation of some biometric characteristics, \eg fixed-length binary vectors for iris, may not be compatible with the fuzzy vault scheme, since it requires sets of integers as input. Therefore, so-called feature type transformations \cite{Lim-BiometricFeatureTypeTransformation-IEEE-2015} have to be applied to the original biometric templates to map them from their representation to an integer set. Ideally, a feature type transformation retains the discriminative information of the biometric features.

\section{Security in Multi-biometric Fuzzy Vaults}\label{sec:considerations}

\subsection{General Security Measures}
The performance of biometric algorithms is usually determined by metrics standardised in ISO/IEC IS 19795-1 \cite{ISO-IEC-19795-1-Framework-210216}. The following metrics are highly relevant for measuring the recognition accuracy of a biometric verification system: 

\begin{itemize}
\item False Non-Match Rate (FNMR): the FNMR is the proportion of completed mated comparison trials that result in a comparison decision of ``non-match''. The Genuine Match Rate (GMR) is defined as GMR$=$1-FNMR.
\item False Match Rate (FMR): the FMR is the proportion of a specified set of completed non-mated comparison trials that result in a comparison decision of ``match''.
\end{itemize}

The FNMR/GMR is commonly interpreted as a measure of usability of the biometric system and the FMR is a measure of security. These performance metrics can be directly applied to measure the biometric performance of a fuzzy vault scheme. In an unprotected biometric system, FNMR/GMR and FMR are adjusted via the decision threshold of the biometric system. In the fuzzy vault scheme, FNMR/GMR and FMR change depending on the parameter configuration of a fuzzy vault, \eg degree of secret polynomial. 

Obviously, there is a relation between the FMR and the security of a fuzzy vault scheme \cite{Veldhuis15}. The False Accept Security (FAS)\footnote{The false accept security is expressed in bits and corresponds to an encryption using a key of that bit-length. From this perspective, the false accept security in bits can be considered as a vague entropy notion.}  provides a good approximation of the security of a fuzzy vault scheme because the false accept attack is typically one of the most efficient attacks.  In a false accept attack, the adversary iteratively simulates non-mated authentication trails until a false match is reached. The FAS (in bits) is often estimated as  $\textrm{FAS}=-\log_2(\textrm{FMR})$. However, it should be noted that an attacker can deviate from the parameters corresponding to the operational FMR, especially in polynomial reconstruction. Additionally, the time required for performing a single authentication attempt should be taken into account when estimating the FAS. 
Many published works report the Brute Force Security (BFS) as a measure of security. Usually, the BFS is derived from the size of the secret key space. However, the BFS often overestimates the effective security of a fuzzy vault scheme. That is, for an attacker it is usually much easier to find a feature set based on which the secret key can be reconstructed from the vault, than guessing the secret key. 

\subsection{Feature Level Fusion}

As mentioned before, fusion in multi-biometric fuzzy vault should be performed at feature level to achieve a high security \cite{Merkle12a}. Assume that the integer sets $\mathbf{A}_1,\dots,\mathbf{A}_N$ have been extracted from $N$ biometric characteristics (or multiple instances of a single biometric characteristic). Furthermore, assume that $\mathbf{B}_1,\dots,\mathbf{B}_N$ are second acquisitions such that  $\mathbf{A}_i$ matches with $\mathbf{B}_i$ for each $i$. Ideally, the integer sets $(\mathbf{A}_1,\dots,\mathbf{A}_N)$ and $(\mathbf{B}_1,\dots,\mathbf{B}_N)$ are fused into new integer sets $\mathbf{A}$ and $\mathbf{B}$, respectively, such that $| \mathbf{A}_1 \cap  \mathbf{B}_1| + \dots + |\mathbf{A}_N \cap \mathbf{B}_N| = |\mathbf{A} \cap \mathbf{B}|$. In other words, all elements are distinct within the fused integer sets. This can be achieved by attaching a unique index to the elements of each of the integers sets that together form the fused integer set. For each $i=1,\dots,N$ we attach the index $i$ to the elements of $\mathbf{A}_i$. Let $x \in \mathbf{A}_i$ denote an integer; then $i +Nx$ may be used as the feature element to which the index $i$ has been attached. Finally, we may use the union of all elements with attached indexes, \ie $\mathbf{A} = \{i+Nx|x\in \mathbf{A}_i, i=1,\dots,N\}$ as the fused integer set. Finally, each integer $x \in \mathbf{A}$ is encoded as element $a$ of a finite field $\mathbf{F}$ of sufficient size resulting in a feature set $\mathbf{P}$. In the same way, the feature set $\mathbf{U}$ is obtained from the integer sets $\mathbf{B}_1,\dots,\mathbf{B}_N$ at the time of verification.

\section{Fuzzy Vault based on Face and Fingerprints}\label{sec:casestudy}

\subsection{Multi-biometric Database}

\begin{figure}[!t]
\vspace{-0.0cm}
\centering
\subfigure[]{\includegraphics[width=0.47\textwidth]{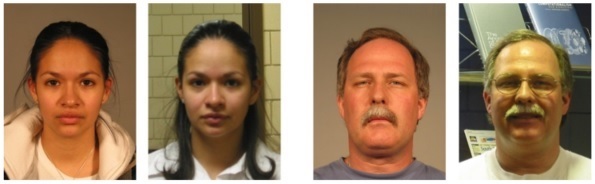}}
\subfigure[]{\includegraphics[width=0.47\textwidth]{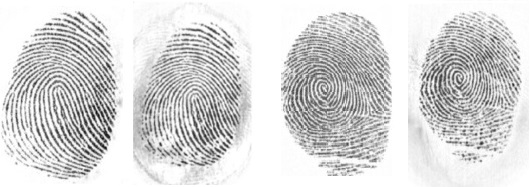}}\vspace{-0.2cm}
\caption{Examples of (a) face and (b) fingerprint image pairs of different subjects of the FRGCv2 and the MCYT-100 databases.}\label{fig:db}\vspace{-0.2cm}
\end{figure}

We construct a multi-biometric system in which the face is fused with four fingerprints. To this end, we use the MCYT-100 fingerprint database \cite{ip-vis_20031078} and a subset of the FRGCv2 face image database \cite{Phillips2005}. The latter subset consists of face images of 533 subjects. Face images are labelled as references (984 images) or probes (1,726 images).  Facial reference images are more constrained (neutral expression, homogeneous illumination etc.) compared to the probe images that are generally less constrained. Similar authentication scenarios may be found in real-world multi-biometric systems, \eg the European Entry/Exit System (EES) \cite{EES}. A different amount of reference and/or probe images may be available for different subjects. As its name indicates, the MCYT-100 fingerprint database contains fingerprints from 100 subjects. Precisely, 12 fingerprint images are available for each of a subject’s ten fingers.  For fingerprint images, there is no expected variation in terms of sample quality between fingerprints captured during enrolment or authentication. Therefore, images can be divided into reference (presented at enrolment) and probe (presented at authentication) fingerprints as required. Examples of reference and probe face images of two subjects are depicted in figure~\ref{fig:db}~(a) and two pairs of fingerprint images are shown in figure~\ref{fig:db}~(b).

In order to maximise the number of virtual subjects in the composed multi-biometric database, the left and right hands of the MCYT-100 are treated as being from different subjects resulting in 200 subjects with 4 fingers (fingerprints of the thumbs are discarded). To further maximize the number of possible comparisons we combine the resulting 200 subjects of the MCYT-100 database with 193 subjects of the FRGCv2 database for which 2 reference and 5 probe images are available and 7 subjects for which 2 reference and 4 probe images are available. This means the first two fingerprint images are assigned to the reference images, respectively, and the next 4 or 5 fingerprints to the probe images, respectively.

The composed database allows performing 1,986 mated comparisons (genuines). For non-mated comparisons, comparisons between fingerprints of the left and right hand of single subjects are excluded\footnote{There may be a certain correlation, e.g. number of minutiae, between fingers of left and right hands of single subjects, which in turn may negatively affect the FMR of the overall system.}. This results in 196,614 non-mated comparisons (impostors).

\subsection{Algorithms}

For the pre-alignment of fingerprints, the xTARP method \cite{Merkle17} is used. We count pre-alignment errors and record these in the Failure-To-Acquire Rate (FTAR) \cite{ISO-Vocabulary-2017}. Minutiae records have been extracted using the FingerJetFX algorithm \cite{fingerjetfx}. For each finger, the minutiae listed in the record are sorted by their quality value, and successively quantised to integers until a predefined upper bound $t_{max}$ is reached. This results in a set of integers of size at most $t_{max}$ per finger. This quantisation process is necessary to protect against correlation attacks \cite{bib:TamsMihailescuMunk2015}. In detail, we use a hexagonal grid of which coordinates  $\Lambda_i$ are equidistantly spaced by $\ell=25$ pixels; the grid is centred in the region in which absolutely pre-aligned minutia can occur. Given a minutia $(\alpha, \beta, \theta)$, its quantization is computed by determining the index $j$ of the grid coordinate  $\Lambda_j$ being closest to the coordinates $(\alpha, \beta)$. The minutia angle $\theta$ is quantised by $s=6$ different quanta encoded by $j'=\lfloor s\theta/2\pi\rfloor$. Finally, the quantisations $j$ and $j'$ of the minutia’s position and orientation, respectively, are combined to $j'+sj$ as the integer encoding of the quantised minutia. 

For face detection, we use the dlib face detector  \cite{King2009}. For feature extraction, we employ the ResNet100 model from \cite{Deng19} published in the InsightFace repository\footnote{\url{https://github.com/deepinsight/insightface/}} as LResNet. Applied to a cropped face image ArcFace extracts face embeddings consisting of 512 float numbers. Float vectors are converted to integer sets following the approach described in \cite{RATHGEB2022102539}: in a first step, the feature space of each float value is divided in four equal-probable intervals, each encoded by three bits using Linear Seperable Subcodes (LSSC) \cite{Lim-LinearlySeparableSubcode-PAMI-2013}; in a second step, an integer set is created from all indexes of 1s in the binary vector.

As decoding algorithm we employ an improved version of the Guruswami-Sudan (GS) list decoding algorithm \cite{bib:GuruswamiSudan1998} published by Trifonov \cite{bib:Trifonov2010}. This list decoder returns a list of candidate polynomials. From these, the correct polynomial can again be determined by checking its hash value. Provided that  $\omega > \sqrt{u (k-1)}$ this algorithm can potentially recover $\kappa$. The computational efficiency can be a traded-off against the number of correctable errors by an additional parameter referred to as multiplicity. The higher the multiplicity, the more errors the algorithm can tolerate while also being more inefficient. For instance, for a multiplicity of 1, the number of correctable errors is approximately $u-\sqrt{2u (k-1)}$. 

\setlength{\tabcolsep}{5pt}
\begin{table}[!t]
\small
\begin{center}
\caption{Performance of the fuzzy vault scheme based on four fingerprints (italic FAS values are interpolated).}\label{tab:finger}\vspace{-0.0cm}
\renewcommand*{\arraystretch}{1.2}
\begin{tabular}{|c|c|c|c|}
\hline
\rowcolor{gray!15} $k$	&	\textbf{GMR (in \%)}	&	\textbf{FMR (in \%)}	&	\textbf{FAS (in bits)}	\\\hline
2	&	100.00	&	97.16345066	&	2.94	\\\hline
4	&	100.00	&	71.18579128	&	4.51	\\\hline
6	&	100.00	&	39.73560511	&	5.86	\\\hline
8	&	100.00	&	19.08081504	&	7.17	\\\hline
10	&	99.90	&	8.274618913	&	8.51	\\\hline
12	&	99.80	&	3.424636399	&	9.77	\\\hline
14	&	99.65	&	1.302174405	&	11.20	\\\hline
16	&	99.29	&	0.466715353	&	12.74	\\\hline
18	&	98.89	&	0.157628932	&	14.29	\\\hline
20	&	98.58	&	0.054000123	&	15.85	\\\hline
22	&	98.18	&	0.019800045	&	17.39	\\\hline
24	&	97.82	&	0.00745716	&	18.73	\\\hline
26	&	97.16	&	0.00077143	&	22.02	\\\hline
28	&	95.85	&	0.000257143	&	23.61	\\\hline
30	&	94.63	&	0	&	\textit{25.20}	\\\hline
\end{tabular}
\end{center}\vspace{-0.2cm}
\end{table}

\subsection{Experiments}

In our experiments, the Failure-to-Acquire rate (FTA) is recorded for the fingerprint pre-alignment step. In case multiple fingerprints are used, a single failure to align will result in a failure to enroll. The Failure-To-Enroll rate (FTE) for the combination of four fingerprint is 1.1\%, due to errors in the pre-alignment process.

Biometric performance is measured in terms of GMR and FMR. Security is measured in terms of FAS. As mentioned earlier, compared to the FAS, the BFS tends to significantly overestimate the effective security of a fuzzy vault scheme \cite{bib:TamsMihailescuMunk2015}. The BFS usually increases with the degree of the secret polynomial. Precisely, for the proposed system the minimum BFS observed (in bits) was similar to the size of the polynomial, \ie $\min(\mbox{BFS}) \approx k$. Therefore, a more realistic measure can be derived from the FMR assuming that the attacker applies the GS decoding. Following the approach in \cite{RATHGEB2022102539}, the FAS is estimated as,
\begin{equation}
t \cdot \log(0.5)/\log(1-\mbox{FMR})
\end{equation}
where $t$ is the average amount of operations for a non-mated verification attempt.  Precisely, the FAS defines the number of operations that an attacker requires to succeed with a probability of 50\% (alternatively, the FAS could be estimated as $t/$FMR, \ie the expected number of steps until an attack succeeds. Note that $t$ depends on the chosen $k$ and is measured in terms of Lagrange interpolation. That is, for the remaining decoding strategies $t$ is measured relatively to the Lagrange interpolation. For some polynomial degrees no false matches have been observed. In such cases, the FAS can not be estimated and is approximated by linearly interpolating the FAS values of the last two polynomial degrees for which the FAS could be estimated.

\setlength{\tabcolsep}{5pt}
\begin{table}[!t]
\small
\begin{center}
\caption{Performance of the fuzzy vault scheme based on the face (italic FAS values are interpolated).}\label{tab:face}\vspace{-0.0cm}
\renewcommand*{\arraystretch}{1.2}
\begin{tabular}{|c|c|c|c|}
\hline
\rowcolor{gray!15} $k$	&	\textbf{GMR (in \%)}	&	\textbf{FMR (in \%)}	&	\textbf{FAS (in bits)}	\\\hline
256	&	100.00	&	62.64737	&	5.79	\\\hline
272	&	100.00	&	37.4190037	&	6.92	\\\hline
288	&	100.00	&	17.6622723	&	8.24	\\\hline
304	&	100.00	&	6.8153845	&	9.75	\\\hline
320	&	100.00	&	2.1951641	&	11.47	\\\hline
336	&	100.00	&	0.6019409	&	13.30	\\\hline
352	&	100.00	&	0.1637727	&	15.20	\\\hline
368	&	99.95	&	0.0467922	&	17.06	\\\hline
384	&	99.90	&	6.10E-03	&	19.98	\\\hline
400	&	99.45	&	1.53E-03	&	21.99	\\\hline
416	&	98.24	&	2.54E-04	&	24.67	\\\hline
432	&	94.11	&	0	&	\textit{27.35}	\\\hline
448	&	87.97	&	0	&	\textit{30.04}	\\\hline
464	&	79.00	&	0	&	\textit{32.72}	\\\hline
480	&	65.11	&	0	&	\textit{35.40}	\\\hline
\end{tabular}\vspace{-0.2cm}
\end{center}
\end{table}

Firstly, the performance of the fuzzy vault based on the individual biometric characteristics is evaluated on the composed database. Obtained results are summarised in  Table~\ref{tab:finger} and Table~\ref{tab:face}, showing the performance of the fuzzy vaults based on four fingerprints and the face, respectively. Due to their differences in terms of feature set sizes and error distribution (see next section), suitable polynomial degrees are of different orders of magnitudes (approx. 300-400 for face and 20-30 for four fingerprints).

Focusing on fingerprints, the results are considerably worse than those reported in the work of Tams \cite{Tams16a}. However, in \cite{Tams16a}, only false accept attacks using Lagrange interpolation were considered. Our analysis shows that this attack is much more efficient using the improved version of the GS algorithm published in \cite{bib:Trifonov2010}. Thus, we conclude that the limitation of the false accept attack to Lagrange interpolation would result in too optimistic security bounds; note that the author of \cite{Tams16a} explicitly warned that it is definitely possible that more efficient attack as those based on mere Lagrange interpolation could be found. With respect to the face, obtained results confirm those reported in \cite{RATHGEB2022102539}, in which the same construction, database, and feature extraction algorithms were used. It was found that the performance rates decrease if only two intervals are used to quantise the feature space of extracted float values, \ie Boolean binarisation.

\setlength{\tabcolsep}{5pt}
\begin{table}[!t]
\small
\begin{center}
\caption{Performance of the multi-biometric fuzzy vault scheme (italic FAS values are interpolated).}\label{tab:fusion}\vspace{-0.0cm}
\renewcommand*{\arraystretch}{1.2}
\begin{tabular}{|c|c|c|c|}
\hline
\rowcolor{gray!15} $k$	&	\textbf{GMR (in \%)}	&	\textbf{FMR (in \%)}	&	\textbf{FAS (in bits)}	\\\hline
256	&	100.00	&	24.4494559	&	7.61	\\\hline
272	&	100.00	&	9.4204501	&	9.16	\\\hline
288	&	100.00	&	2.8810352	&	10.97	\\\hline
304	&	100.00	&	0.6945444	&	13.09	\\\hline
320	&	100.00	&	0.1530003	&	15.33	\\\hline
336	&	100.00	&	0.0308572	&	17.59	\\\hline
352	&	100.00	&	3.34E-03	&	20.81	\\\hline
368	&	100.00	&	0	&	\textit{24.03}	\\\hline
384	&	100.00	&	0	&	\textit{27.26}	\\\hline
400	&	100.00	&	0	&	\textit{30.48}	\\\hline
416	&	99.90	&	0	&	\textit{33.70}	\\\hline
432	&	99.39	&	0	&	\textit{36.92}	\\\hline
448	&	97.77	&	0	&	\textit{40.14}	\\\hline
464	&	94.93	&	0	&	\textit{43.37}	\\\hline
480	&	90.07	&	0	&	\textit{46.59}	\\\hline
\end{tabular}\vspace{-0.2cm}
\end{center}
\end{table}

\begin{figure}[!t]
\vspace{-0.0cm}
\centering
\includegraphics[width=0.45\textwidth]{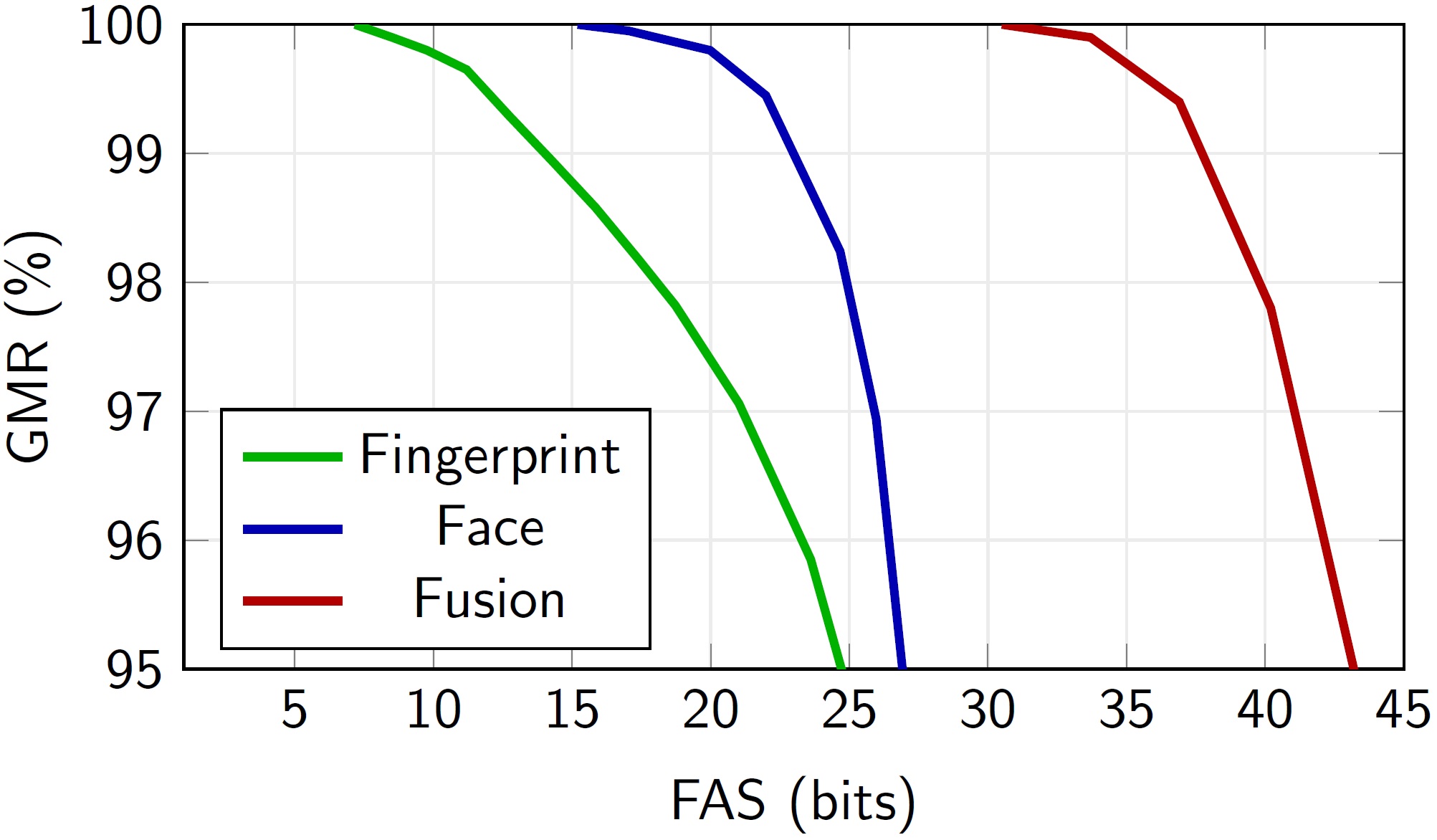}
\caption{Relation between the GMR and FAS in the fuzzy vaults.}\label{fig:gmrfas}\vspace{-0.0cm}
\end{figure}

Table~\ref{tab:fusion} summarises the performance rates obtained for multi-biometric fuzzy vault. It can be seen that the performance rates significantly improve compared to the individual fuzzy vault schemes based on the face or fingerprints alone. For the fusion of four fingerprints and the face, an GMR of 100\% and FMR of 0\% are obtained, \eg at a polynomial degree of 400 – this means the multi-biometric fuzzy vault provides a perfect recognition accuracy on the composed multi-biometric database. Compared to the individual fuzzy vault schemes based on a single biometric characteristic, the security (in terms of FMR) as well as usability (in terms of GMR) is improved. The relation between the GMR and the FAS for the fuzzy vault schemes based on the face and four fingerprints as well as the multi-biometric fuzzy vault is illustrated in figure~\ref{fig:gmrfas}. Note that shown values must be treated with care, since many data points are extrapolated, especially for the multi-biometric fuzzy vault.

\section{Discussion}\label{sec:discussion}
\subsection{Feature Balancing}
While some works have proposed constructions of multi-biometric fuzzy vaults similar to the one presented in this work, they commonly ignore certain issues that may negatively impact the overall system's security. In case a fusion of different biometric characteristics is performed, two important issues are expected to arise\footnote{These issues are expected to occur less in case features to be fused are obtained from different instances of a single biometric characteristic, \eg multiple fingerprints or two irises of the same subject.}:
\begin{enumerate}
\item	\textit{Error distributions}: integer sets may exhibit varying biometric variances (intra- and inter-class). This has been shown for fuzzy vault schemes based on different biometric characteristics. For instance, for non-mated comparisons, integer sets extracted from deep face representation are expected to overlap in at least 50\% \cite{RATHGEB2022102539} while for non-mated comparisons of minutiae sets usually only up to 20\% of minutiae are expected to match \cite{Tams16a}. This means, an attacker could focus on launching a false accept attack on a single characteristic for which it is known that non-mated comparisons result in a large number of matching elements.
\item \textit{Set sizes}: it is likely that corresponding integer sets are of different size. In a fusion, an implicit weighting is introduced according to the set sizes, \ie larger sets are stronger weighted and vice versa. Such imbalance can in turn have a negative impact on the overall security of a multi-biometric fuzzy vault. Again, an attacker could merely focus on launching a false accept attack on the biometric characteristic, which constitutes the largest relative part of the fused template. 
\end{enumerate}

\begin{table}[!t]
\begin{center}
\caption{Average set sizes and overlapping elements for mated and non-mated integer sets.}\label{tab:stats}\vspace{-0.0cm}
\small
\renewcommand*{\arraystretch}{1.2}
\begin{tabular}{|c|c|c|c|}
\hline
\rowcolor{gray!15}& \textbf{Average} & \multicolumn{2}{c|}{\textbf{Average Overlap}} \\
\rowcolor{gray!15}\textbf{Characteristic}  & \textbf{Set size} & \textbf{\quad Mated \quad} & \textbf{Non-Mated} \\\hline
Face & 769.83 & 615.04 & 449.40 \\ \hline
Four Fingerprints & 161.45 & 89.92 & 21.48 \\\hline
Fusion & 931.28 & 704.96 & 470.88 \\\hline
\end{tabular}
\end{center}\vspace{-0.2cm}
\end{table}

For the feature vectors of the different characteristics used in the multi-biometric fuzzy vault of this work, we analyse their size as well as error distributions. Table~\ref{tab:stats} lists the average size of integer sets as well as the average overlap. The overlaps between mated and non-mated comparisons refer to the intersection of the corresponding feature sets, \ie $|\mathbf{A} \cap \mathbf{B}|$, where $\mathbf{A}$ and $\mathbf{B}$ are from the same or different subjects, respectively. Based on the above table we make the following observations:
\begin{enumerate}
\item Facial integer sets are larger than fingerprint integer sets (more than four times); 
\item The relative overlap for non-mated authentications is higher for facial integer sets (more than 20 times).
\end{enumerate}

These observations imply that generally more weight is assigned to the face feature vector. As mentioned before, this means that an attacker could increase his success chance by launching attacks only on the face. From tables~\ref{tab:face} and \ref{tab:fusion}, it can be observed that for a polynomial degree that is optimal for the multi-biometric fuzzy vault, \ie $k=$400, the GMR for using only the face is 99.45\%. This means if the attacker is able to guess or spoof the face features, he has very high chances to unlock the vault which would also allow a reconstruction of the fingerprint features. Even for larger polynomial degree, \eg 480, the success chance for unlocking the multi-biometric fuzzy vault with only the facial features is still above 65\%. This observation indicates that the use of fingerprints in the multi-biometric fuzzy vaults primarily improves the GMR while the security level would be similar to that of the face-based fuzzy vault. 

Due to the above listed issues, it is preferable to \emph{balance integer sets} in a way that these exhibit similar properties in terms of error distribution and set size. This can be achieved in a 2-step process where in the first step, the relative overlap is adjusted, and in the second step, the size of the feature sets (without changing the relative overlap again).

We assume that a function $f: \mathbf{A}_i \rightarrow f(\mathbf{A}_i)$ is available which makes it possible, for a distinct feature extractor of biometric characteristic, to increase the expected relative number of overlapping elements in non-mated comparisons. Let $\omega_i=E(2|\mathbf{A}_i \cap \mathbf{B}_i|/(|\mathbf{A}_i| + |\mathbf{B}_i|))$ and $\omega_j = E(2|\mathbf{A}_j \cap \mathbf{B}_j|/(|\mathbf{A}_j| + |\mathbf{B}_j|))$ denote the expected relative number of overlapping elements for performing a non-mated authentication attempt, \ie $\mathbf{A}$ and $\mathbf{B}$ are obtained from different subjects. We can assume, without loss of generality, $\omega_i<\omega_j$. The function $f$ should artificially increase the relative number of matching elements for non-mated comparisons of $f(\mathbf{A}_i)$ such that it is similar to $\omega_j$. It may require to directly adapt the biometric feature extractor and is thus dependent of the biometric characteristic. It can be assumed that  for performing non-mated comparison trails, the relative amount of matching elements can be increased by applying strong quantisation during feature extraction.  After applying $f$ in the first step, it is likely that $|f(\mathbf{A}_i)| \neq |\mathbf{A}_j|$, \ie the feature set sizes need to be adjusted. In this case, it is possible to increase the size of either integer set in the second step.

We assume that there is a function $g: \mathbf{A}_i \rightarrow g(\mathbf{A}_i)$ which allows to increase the size of an integer set. Let $l_i = |\mathbf{A}_i|$ and $l_j= |\mathbf{A}_j|$, $i\neq j$, denote the sizes of two integer sets to be fused in a fuzzy vault.  Again, for integer sets obtained from different biometric characteristics, without  loss of generality, we can assume that $l_i< l_j$, which has the previously discussed security implications. This means the integer set $\mathbf{A}_i$ should be  artificially increased. This can be achieved by applying a function $g$ that clones each element $x_i \in \mathbf{A}_i$ until $|g(\mathbf{A}_i)| \approx l_j$. Precisely, each element $x_i$ can be cloned $m=\lceil l_j /l_i\rceil$ times by iterating through the residual classes smaller than $m$ and adding each residual $r \in  \mathbb{Z}_{m-1}$ to $x_i m$ to create a new set element $x'_{i,r}=x_i m + r$. 

Using functions $f$ and $g$ as discussed above, two feature sets can be balanced for feature level fusion as follows: First, the expected relative overlap and the average size are empirically determined. Subsequently, $f$ can be applied to the integer set that exhibits a lower relative amount of matching elements in non-mated authentication attempts in order to adjust it to the other one. Finally, $g$ can be applied to adjust the size of the smaller feature set to that of the other one. Note that the application of $g$ does not affect the relative overlap. 
As a consequence, both feature sets have the same expected overlap and the same average size. In case of more than two feature sets the procedure can be iteratively applied by first determining the largest overlap/ size and subsequently increasing the overlaps/ sizes of all other sets to largest one(s).  Thereby, the chances of an attack focusing on a single biometric characteristic would be significantly decreased.    

In order to apply this feature balancing approach to our fuzzy vault construction, we tried to increase the relative number
of matching elements for non-mated comparisons by using stronger quantisations for the minutiae data. More precisely, we increased the values of $\ell$, \ie the pixel distance between points defining the hexagonal grid for minutiae quantisation. While this increased the relative overlap of quantised minutiae for fingerprint comparisons, we observed a drastic decrease of biometric performance. Therefore, we conclude that the feature balancing is not successful on our fuzzy vault. The details of this evaluation are omitted in this work.


Another possible way of enhancing the security in a multi-biometric fuzzy vault would be to harmonize the feature extraction methods. In particular, deep neural networks have been applied to extract fixed-length float vectors from various biometric characteristics \cite{Sundararajan-DeepLearningBiometrics-2018} including fingerprints \cite{Engelsma21}. Such deep features can be efficiently transformed to integer sets in the was as the face features in our multi-biometric fuzzy vault. The use of such a common feature extraction method may enable a fusion of integer sets that already exhibit similar properties.

\subsection{Extension with Password}

A (multi-biometric) fuzzy vault can be further extended with a password or PIN in order to improve its security. In contrast to the biometric features, the password is required to match perfectly to unlock the fuzzy vault. The main idea behind this so-called “password-hardening” is that the security of the fuzzy vault (in bits) is increased by the entropy of the added password. This has firstly been suggested by Nandakumar \etal \cite{Nandakumar07}, who suggest to apply a random transformation function derived from the user password to the biometric template. The transformed template is then secured using the fuzzy vault. Finally, the vault is encrypted using a key derived from the password. The same concept is also applied in \cite{MEENAKSHI2010195}. A similar approach without transforming the features prior to encrypting the fuzzy vault with standard symmetric encryption is suggested by Tams \cite{Tams16a}. 

For extending the multi-biometric fuzzy vault with a password $K$ we propose the following procedure: let  $(\mathit{enc})_K,(\mathit{dec})_K:{0,1}^L \rightarrow{0,1}^L$ denote the symmetric encryption and its decryption functions, respectively of block length $L$ such that $(\mathit{enc})_K  \circ (\mathit{dec})_K=\mathit{id}$, where $\mathit{id}$ denotes the identity operator. That is, the function $id:M\rightarrow M$ such that $id(x)=x$ for all $x\in M$. The operator $\circ$ denotes the concatenation operator. That is, for two functions $f:E\rightarrow D$ and $g:D \rightarrow E$, the expression $h=f \circ g$ denotes the function $h:M\rightarrow M$ such that $h(x)=f(g(x))$ for all $x\in D$. Note, that the functions $(enc)_K$ and $(dec)_K$ typically employ a key derivation function (KDF) that generates a symmetric key from a password/passphrase eligible for being used in a symmetric cipher (\eg AES). 

In our implementation we use a simple mechanism to generate an AES-128 key from a password or passphrase. First, from a password/passphrase, the SHA-1 hash value is computed, and from the resulting 160 bits, the first 128 bits are used as the AES-128 key. For the vault polynomial $V(X)$ we write, $V(X)=X^t+\sum_{j=0}^{t-1}v_j X^j$. We assume, without the loss of generality, that $v_j$ are the integer encodings from the binary finite field $\mathbf{F}$ with at most $e$ bits. Let $v_0 |v_1 |…|v_{t-1} \in\{0,1\}^{te}$
be the concatenation of the bits encoding the vault polynomial $V(X)$. Note, the leading coefficient of $V(X)$ does not need to be considered explicitly, since it is equals to 1 for $k<t$.
To optionally encrypt the fuzzy vault $V(X)$ by using $enc_K$, one can encrypt the bits $v_0 |v_1 |…|v_{t-1}$ padded with random bits such that a multiple of the block length $L$ is reached in combination with an arbitrary block cipher mode (\eg cipher block chaining mode). The encrypted bits $E \in\{0,1\}^{\lceil te/L\rceil L}$ will be stored instead of the vault polynomial and along with the parameters $t$ to allow safe decryption.

To decrypt an encrypted vault $(E,t)$ using a password $K’$, the decrypted bits $D \in\{0,1\}^{\lceil te/L \rceil L}$ are computed first by applying $dec_K'$ to $E$ in the chosen block cipher mode. Then only the first $te$ bits of $D$ are considered (the last $\lceil te/L\rceil  L-te$ are dismissed). It is easy to see that if $K=K’$, then the first $te$ agree with $v_0 |v_1 |…|v_{t-1}$ and the polynomial $V(X)$ can be recovered using the formula $V(X)=X^t+\sum_{j=0}^{t-1}v_j X^j $. Otherwise, if $K'\neq K$, then a spurious polynomial $V’(X)$ will be decrypted. Since we can neglect the probability that decryption with a wrong password results in fuzzy vault by the correct secret polynomial, the entropy of the password fully contributes to the security: The false-accept security of the fuzzy vault is given by the sum of the entropy of the password and the entropy of the biometric data. 

Depending on the length of the password, the resulting two-factor authentication scheme would provide an increased security; not only compared to the sole use of multiple biometrics but also compared to the individual use of multiple biometric characteristics and a password. However, it is important to note that the additional requirement of a password is expected to negatively affect the usability of the system since passwords have to be remembered by users.



\section{Conclusion}\label{sec:conclusion}
In this work, it was shown that the biometric recognition performance of a fuzzy vault improves when feature vectors obtained from the face and fingerprints are concatenated at feature level. Fusion-based performance gains mainly improve the usability (associated with a higher GMR) while also enhancing the security (associated with a lower FMR). At the same time, it we also discussed that the security may be impaired by feature vector imbalance. Feature vectors extracted from different biometric characteristics are likely to differ in size. Moreover, the error distributions yielded by comparisons of feature vectors of different biometric characteristics are expected to be different. Countermeasures to tackle these issues are theoretically discussed in this work. However, the investigation of suitable methods for balancing feature sets in  multi-biometric cryptosystems represent an open challenge and may be subject to future work.

\bibliographystyle{IEEEtran}
\bibliography{references}

\end{document}